\documentclass[intlimits,twoside,a4paper]{article}

\usepackage[greek,english]{babel}
\RequirePackage{color}
\definecolor{MyDarkGreen}{rgb}{0.1,0.60,0.2}

\usepackage{amsmath,amssymb}
\usepackage{graphicx}

\usepackage[eqsecnum]{cmpj2}
\usepackage{teubner}

\def\qq{\hbox{\fontencoding{LGR}\fontfamily{mtr}\selectfont\foreignlanguage{greek}{\coppa}}}
\def\sqq{{\hbox{\fontencoding{LGR}\fontfamily{mtr}\selectfont\foreignlanguage{greek}{\footnotesize\coppa}}}}
\def\tqq{{\hbox{\fontencoding{LGR}\fontfamily{mtr}\selectfont\foreignlanguage{greek}{\tiny\coppa}}}}

\articletype{Regular article}

\title[New critical exponents]%
{A new critical exponent $\qq$ and its logarithmic counterpart $\hat{\qq}$
}
\author[Kenna \& Berche]{Ralph Kenna\refaddr{label1} and
        Bertrand Berche\refaddr{label2},}

\addresses{
\addr{label1} Applied Mathematics Research Centre,
              Coventry University,
              Coventry, CV1 5FB, England
\addr{label2} Statistical Physics Group, Institut Jean Lamour, UMR CNRS 7198,
Universit{\'{e}} de Lorraine, 
B.P. 70239, 54506 Vand\oe uvre l\`es Nancy Cedex, France
}
\begin{document}

\maketitle

\begin{abstract}
It is well known that standard hyperscaling breaks down above the upper critical dimension $d_c$, where the critical exponents take on their Landau values. 
Here we show that this is because, in standard formulations in the thermodynamic limit, distance is measured on the correlation-length scale. 
However, the correlation-length scale and the underlying length scale of the system are not the same at or above the upper critical dimension. 
Above $d_c$ they are related algebraically through a new critical exponent $\qq$, while at $d_c$ they differ through logarithmic corrections governed by an exponent $\hat{\qq}$.
Taking proper account of these different length scales allows one to extend hyperscaling to all dimensions.
\pacs 64.60.-i,64.60.an,05.50.+q,64.60.De,11.10.Kk
\end{abstract}

\section{Introduction}
\label{Introduction}

Since the 1960's, the scaling relations between critical exponents have been of fundamental importance in the theory of critical phenomena \cite{FiHa83,history,Stanley}.
Six primary critical exponents, $\alpha$, $\beta$, $\gamma$, $\delta$, $\eta$ and $\nu$, have played the most important roles and these are related by four famous scaling relations. 
One of these --- the  hyperscaling relation --- involves the dimensionality $d$ of the system. 
It has long been known that hyperscaling, in its standard form, fails above the upper critical dimension $d=d_c$ where the critical exponents take their Landau, mean-field values. 
E.g.,  for the Ising model  above $d_c=4$, one has $\alpha = 0$, $\beta = 1/2$, $\gamma = 1$, $\delta = 3$, $\eta = 0$ and $\nu = 1/2$, irrespective of the dimensionality $d$.
Here we report on a  more complete form for the hyperscaling relation which holds in all dimensions \cite{BeKe12}. 
This involves a new critical exponent which we denote by ${\qq}$ (pronounced ``koppa'' \cite{coppa}) and which characterises the finite-size scaling (FSS) of the correlation length. We report evidence for the universality of ${\qq}$ through numerical studies of the five-dimensional Ising model with  free  boundary conditions \cite{BeKe12}. 

We also examine hyperscaling {\emph{at}} the upper critical dimension, which is characterised by  multiplicative logarithmic corrections. 
These corrections are also characterised by critical exponents which  have scaling relations between them. 
The logarithmic hyperscaling relation involves an exponent ${\hat{\qq}}$ which is the logarithmic analogue of ${\qq}$ \cite{KJJ2006}.

We consider a lattice spin system in $d$ dimensions.  
In units of the lattice constant, its linear extent is $L$. 
We denote by $P_L(t)$ the value of a function $P$ measured on such a system at reduced temperature $t$. The latter is defined as 
\begin{equation}
 t = \frac{T-T_L}{T_L},
\label{t}
\end{equation}
where $T_L$ is the value of the temperature $T$ at which the finite-size reduced susceptibility (defined below) peaks and is refered to as the pseudocritical point. 

In the infinite-volume limit, $T_L$ becomes the critical point $T_\infty \equiv T_c$
and the specific heat and correlation length scale nearby as
\begin{equation}
 c_\infty(t) \sim t^{-\alpha},
 \quad
 \xi_\infty(t) \sim t^{-\nu}.
 \label{01}
\end{equation}
The standard form of the hyperscaling relation, which is valid at and below the upper critial dimension, links the critical exponents in equation (\ref{01}), 
\begin{equation}
 \nu d = 2 - \alpha.
 \label{hyperscaling}
\end{equation}
Equation (\ref{hyperscaling}) was proposed by Widom \cite{Wi65}.
Kadanoff later presented an alternative but similar argument for it \cite{Ka66}
and Josephson derived the related inequality $\nu d \ge 2 - \alpha$
on basis of plausible but non-rigorous assumptions \cite{Jo67}.
The scaling relations, including equation (\ref{hyperscaling}), are now well understood through the renormalization group \cite{Ma}.

The Landau or mean-field values $\alpha = 0$ and $\nu = 1/2$ for the Ising model  are well established for all values of $d$ at and above the upper critical dimension $d_c=4$.
Since $\alpha$ and $\nu$ are fixed for  $d > d_c$, equation (\ref{hyperscaling}) cannot hold there.
This is referred to as the collapse of hyperscaling in high dimensions.

Here we introduce a new critical exponent ${\qq}$ which characterises the leading FSS of the correlation length,
\begin{equation}
 \xi_L(0) \sim L^{\sqq}.
 \label{sqq}
\end{equation}
We show that the incorporation of $\qq$ into the hyperscaling relation (\ref{hyperscaling}) via
\begin{equation}
 \frac{\nu d}{\qq} = 2 - \alpha,
 \label{hhyperscaling}
\end{equation}
renders it valid in all dimensions (with $\qq=1$ in $d \le d_c$ dimensions).

The critical dimension itself is characterised by multiplicative logarithmic corrections, so that equations (\ref{01}) and (\ref{sqq}) become
\begin{equation}
 c_\infty(t) \sim t^{-\alpha} |\ln{t}|^{\hat{\alpha}},
 \quad
 \xi_\infty(t) \sim t^{-\nu} |\ln{t}|^{\hat{\nu}},
 \label{01log}
\end{equation}
 and
 \begin{equation}
 \xi_L(0) \sim L (\ln{L})^{\hat{\sqq}},
 \label{sqqhat}
\end{equation}
at $d=d_c$, respectively.
Here we also show that the logarithmic analogue of the hyperscaling relation at the upper critical dimension is
\begin{equation}
  \hat{\alpha}  
  =  
  d (\hat{\qq} -  \hat{\nu}) .
  \label{SRlog1}  
\end{equation}
Caution: this last scaling relation has been shown to hold {\emph{at}} the upper critical dimension in a variety of models including the Ising and $O(N)$ $\phi^4$ models, their counterparts with long-range interactions, $m$-component spin glasses, the percolation and Yang-Lee edge problems. 
It does {\emph{not}} hold in {\emph{some}} cases of logarithmic corrections {\emph{below}} the upper critical dimension when the leading exponent $\alpha$ vanishes. 
In these anomalous circumstances, an extra multiplicative logarithmic correction  appears, as explained below and in reference \cite{KJJ2006}.
E.g., the pure Ising model in two dimensions has $\hat{\qq}  =   \hat{\nu} = 0$ but $\hat{\alpha} = 1$.
The random-site or random-bond Ising model  in $d=2$ has $\hat{\qq}  = 0$, $\hat{\nu} = 1/2$ but $\hat{\alpha} = 0$.
The reason for the extra logarithm in these cases is well understood and briefly given in Section~\ref{Derivation}.
Here we are only concerned with $d \ge d_c$, so we refer the reader to reference \cite{KJJ2006} for details of these anomalous  cases in $d<d_c$ dimensions.

The $d>d_c$ hyperscaling relation (\ref{hhyperscaling}) was derived in reference \cite{BeKe12}
and its logarithmic counterpart (\ref{SRlog1}) was developed in references \cite{KJJ2006}.
Next, both of these derivations are summarised.

\section{Derivation of the new hyperscaling relations}
\label{Derivation}

We begin with more general forms for the scaling of the susceptibility and correlation length in infinite volume, encompassing leading behaviour both at and above the upper critical dimension in the thermodynamic limit, namely
\begin{equation}
 \chi_\infty \sim t^{-\gamma} (\ln{t})^{\hat{\gamma}},
 \quad \quad
 \xi_\infty \sim t^{-\nu} (\ln{t})^{\hat{\nu}}.
\label{02}
\end{equation}
The  derivation which we are about to present involves a type of self-consistency analysis using the zeros of the partition function. 
The Lee-Yang zeros are those points in the complex $h$-plane at which the partition function $Z_L(t,h)$ vanishes \cite{LY}. 
Under very general conditions, Lee and Yang proved these to be located on the imaginary $h$-axis, although this is not a pre-requisite for what is to follow here. 
What is required, however, is the notion of the so-called Yang-Lee edge. In the infinite-volume limit, this is the end point of the distribution of zeros which lies closest to the real $h$-axis. 
As such, it most strongly influences critical behaviour. 
In line with the above ans{\"{a}}tze, we assume that the Yang-Lee edge scales as 
\begin{equation}
 h_{\rm{YL}}(t) \sim t^{\Delta} (\ln{t})^{\hat{\Delta}}.
\label{03}
\end{equation}
Here, $\Delta$ is the gap exponent and $\hat{\Delta}$ is its logarithmic counterpart. These are given through static scaling relations \cite{KJJ2006}
\begin{equation}
 \alpha = 2 + \gamma - 2\Delta , \quad \quad \hat{\alpha} = \hat{\gamma} + 2 \hat{\Delta} .
\label{04}
\end{equation}
Our final ingredient is to promote equation (\ref{sqqhat}) to the more general form, 
\begin{equation}
 \xi_L(0) \sim L^{\sqq} (\ln{L})^{\hat{\sqq}}.
 \label{05}
\end{equation}
In each of equations (\ref{02})-(\ref{05}), the hatted exponents vanish above the upper critical dimension \cite{Butera}.
They play an important role {\emph{at}} $d_c$ itself. 
Circumstances in which they are non-vanishing below the upper critical dimension are not our main concern here.  

We write the finite-size partition function in terms of the Lee-Yang zeros $h_j$ as
\begin{equation}
 Z_L(t,h) = 
 A
 \prod_{j=1}^{L^d}{(h-h_j(t,L))}.
\label{06}
\end{equation}
Here the zeros $h_j$, which are dependent on both $t$ and $L$, are ordered such that the smaller the index $j$, the closer the zero is to the real $h$-axis. In this way, $h_1$ is the finite-size counterpart to the Yang-Lee edge. 
The prefactor $A$ plays no important role in what is to come and we henceforth drop it.
The reduced free energy per unit volume is
\begin{equation}
 f_L(t,h) = -L^{-d} \ln{Z_L(t,h)} \sim -L^{-d}\sum_{i=1}^{L^d}{\ln{(h-h_j(t,L))}}.
\label{07}
\end{equation}
Differentiating twice with respect to field delivers the (reduced) finite-size susceptibility as
\begin{equation}
 \chi_L(t,h) \sim \frac{\partial^2 f_L(t,h)}{\partial h^2} = \frac{1}{L^d} \sum_{j=1}^{L^d}{\frac{1}{(h-h_j(t,L))^2}}.
\label{08}
\end{equation}
From now on, we set $h=t=0$ and drop the corresponding  arguments in $\chi_L$ and $h_j$. 
The finite-size pseudocritical susceptibility is therefore
\begin{equation}
 \chi_L \sim \frac{1}{L^d}\sum_{j=1}^{L^d}{\frac{1}{h_j^2(L)}}.
\label{09}
\end{equation}
Equation (\ref{09}) constitutes the consistency check we have been looking for;
it relates the finite-size susceptibility to the $L$-dependency of the Lee-Yang zeros.

FSS is obtained by replacing infinite-volume correlation length in equation (\ref{02})
by the finite-size expression  (\ref{05}), $\xi_\infty \rightarrow  \xi_L$. 
This means that FSS amounts to replacing
\begin{equation}
 t \rightarrow L^{-\frac{\tqq}{\nu}}(\ln{t})^{\frac{\hat{\nu}-\hat{\tqq }}{\nu}}. 
\label{10}
\end{equation}
The susceptibility in equation (\ref{02}) and the edge in equation (\ref{03}) then take the FSS forms
\begin{equation}
 \chi_L \sim L^{\frac{\tqq\gamma}{\nu}}(\ln{L})^{\hat{\gamma}- \gamma \frac{\hat{\nu}-\hat{\tqq}}{\nu}}, 
 \quad \quad
 h_1(L) \sim L^{-\frac{\tqq\Delta}{\nu}}(\ln{L})^{\hat{\Delta}+ \Delta \frac{\hat{\nu}-\hat{\tqq}}{\nu}},
\label{11}
\end{equation}
In fact, following reference \cite{IPZ}, one expects the higher-index zeros to scale as a function of a fraction of the total number of zeros, $j/L^d$.
This expectation allows us to promote the second formula in equation (\ref{11}) to
\begin{equation}
 h_j(L) \sim \left({\frac{j}{L^d}}\right)^{\frac{\tqq\Delta}{\nu d}}\left({\ln{\left({\frac{j}{L^d}}\right)}}\right)^{\hat{\Delta}+ \Delta \frac{\hat{\nu}-\hat{\tqq}}{\nu}}.
\label{12}
\end{equation}
Inserting equations (\ref{11}) and (\ref{12}) into the consistency expression (\ref{09}), one finds
\begin{equation}
 L^{\frac{\tqq\gamma}{\nu}} (\ln{L})^{\hat{\gamma} - \gamma \frac{\hat{\nu}-\hat{\tqq}}{\nu}}
 \sim 
 L^{\frac{2 \tqq \Delta}{\nu}-d}
 \left({\ln{L}}\right)^{-2\Delta \frac{\hat{\nu}-\hat{\tqq}}{\nu}-2\hat{\Delta}}
 \sum_{j=1}^{L^d}{\left({\frac{1}{j}}\right)^{2\frac{\tqq\Delta}{\nu d}}}.
\label{13}
\end{equation}
This is the equation from which we will now draw the hyperscaling relations at and above $d_c$.

Firstly we assume that there are, in fact, no leading logarithmic corrections, so that 
$\hat{\gamma}=0$, $\hat{\nu}=0$, $\hat{\Delta}=0$, $\hat{\qq}=0$. 
This is the circumstance above the upper critical dimension as recently confirmed by Butera and Pernici \cite{Butera}.
Even without these logarithmic corrections, the sum on the right hand side of equation (\ref{13}) generates an extra logarithm if $2q\Delta = \nu d$. 
In the Ising case above $d=d_c$, mean-field theory gives $\nu = 1/2$ and $\Delta = 3/2$.
If $\qq=1$ there, one has logarithmic corrections in $d=6$.
This is a contradiction to the results established in reference \cite{Butera}. Therefore $\qq$ cannot, in fact, be $1$ above $d=d_c$.

This is an important result. 
It means that the finite-size correlation length is {\emph{not}} comensurate with the length above $d=d_c$.  This is contrary to explicit statements in reference \cite{BNPY} and other literature. 
The standard form for hyperscaling (\ref{hyperscaling}) only holds when the two length scales coincide.
Above the upper critical dimension, one must account for the fact that they differ. 
Our main result is derived by equating the leading power-laws for $L$ in equation (\ref{13}). 
Inputting the leading static scaling relation in equation (\ref{04}) then delivers the new expression for hyperscaling in equation (\ref{hhyperscaling}).

The standard leading hyperscaling relation (\ref{hyperscaling}) is valid in $d=d_c$ dimensions so that $\nu d_c = 2-\alpha$.
Combining this with the new expression (\ref{hhyperscaling}) leads to 
\begin{equation}
 {\qq}  =  \frac{d}{d_c} ,
 \label{q}
\end{equation}
for $d \ge d_c$. Since equation (\ref{hyperscaling}) is also valid in $d<d_c$ dimensions, one has, from equation (\ref{hhyperscaling}) that $\qq=1$ there.
The result (\ref{q}) for $d \ge d_c$ agrees with an explicit analytical calculation by Br\'ezin \cite{Br82} for the large-$N$ limit of the $N$-vector model with periodic boundary conditions (PBC's).

However, boundary conditions do not play a role in our derivation of equation (\ref{hhyperscaling}).
This means, in particular, that the correlation length exceeds the actual length of the system close to $t=0$ for free boundary conditions (FBC's) as well as for PBC's. 
Again, this is contrary to many statements in the literature \cite{BNPY,RuGa85,JoYo05,LuMa11}.
In Section~\ref{Numerical}, we will verify numerically that equation (\ref{q}) indeed holds for FBC's.

The logarithmic counterpart of the hyperscaling relation comes from equating powers of logarithms in (\ref{13}).
Inserting the static relation for the correction exponents from equation (\ref{04}) then delivers equation (\ref{SRlog1}) when $d=d_c$.

As stated in the Introduction, the specific heat takes an extra  logarithmic correction, beyond that coming from the hyperscaling relation (\ref{SRlog1}), below the upper critical dimension in special circumstances. 
These circumstances involve the impact angle  $\phi$ at which the complex-temperature (Fisher) zeros impact onto the real axis in the thermodynamic limit.  If $\alpha = 0$, and if this impact angle\index{impact angle} is any value other than $\pi/4$, an extra logarithm arises in the specific heat.
This happens in $d=2$ dimensions for example, but not in $d=4$, where $\phi = \pi/4$ \cite{IPZ}. The reader is referred to reference \cite{KJJ2006} for details of this anomaly.

The logarithmic critical exponent $\hat{\qq}$ was originally introduced as $\hat{q}$ in reference {\cite{KJJ2006} along with the scaling relation (\ref{SRlog1}).
That scaling relation (and its anomalous counterpart, in some $d<d_c$ cases) was tested, and verified, against known results in the literature there. 
The fact that the other exponents in equation (\ref{SRlog1}) are universal indicates that  $\hat{\qq}$ is universal too.

Equation (\ref{hhyperscaling}) is the leading-scaling counterpart of the logarithmic hyperscaling relation (\ref{SRlog1}).
Our claim in this paper is that $\qq$ is a new critical exponent which is universal --- as are the other exponents appearing in equation (\ref{hhyperscaling}). 
The extended hyperscaling relation (\ref{hhyperscaling}) or (\ref{q}) is therefore of similar status to the other standard scaling relations.
Like them, equation (\ref{hhyperscaling}) holds in all dimensions, including $d>d_c$.

The leading part of equation (\ref{11}) gives the leading FSS of the susceptibility and the Yang-Lee edge at the pseudocritical point $t=0$. If $\qq$ is universal as we claim, FSS should also be universal above $d_c$.
However, the standard belief in the literature is that above the upper critical dimension, FSS is {\emph{not}} universal \cite{BNPY,RuGa85,JoYo05,LuMa11}.
In particular, until now, FSS with FBC's has been believed to differ from FSS with periodic boundaries.
We next explain how our theory differs from standard literature regarding FSS and we then go on to provide evidence which (a) idenifies ours as correct and (b) explains  the difference between them.

\section{Finite-size scaling}
\label{Finite-size}

Equation (\ref{11}) gives the FSS of the susceptibility and the Yang-Lee edge at the pseudocritical point $t=0$. 
In these formulae, $\gamma$, $\nu$ and $\Delta$ assume their mean-field values $1$, $1/2$ and $3/2$, respectively, for $d \ge d_c$.

Setting $\qq=1$ gives the FSS {\emph{at}} the critical dimension.
For the $O(N)$ model with $\hat{\gamma} = (n+2)/(n+8)$, $\hat{\nu} = (n+2)/2(n+8)$, $\hat{\Delta}=(1-n)/(n+8)$ and $\hat{\qq}=1/4$
\cite{Br82,BLZ,KeLa91,Ke04}, one obtains
\begin{equation}
 \chi_L(0) \sim L^{2} (\ln{L})^{1/2}, \quad \quad h_1(L) \sim L^{-3} (\ln{L})^{1/4},
\label{100}
\end{equation}
independent of $n$, a result already derived in reference \cite{Ke04} and verified numerically in the periodic Ising case in references \cite{KeLa91,KeLa93,JoYo05}. 

Above the upper critical dimension, where there are no leading multiplicative logarithmic corrections \cite{Butera}, FSS for the susceptibility and edge is given by
\begin{equation}
 \chi_L(0) \sim L^{\frac{\tqq\gamma}{\nu}}, \quad \quad h_1(L) \sim L^{\frac{\tqq\Delta}{\nu}},
\label{101}
\end{equation}
respectively.
If it were the case that $\qq=1$,  equation (\ref{101}) would reduce to 
\begin{equation}
 \chi_L(0) \sim L^{\frac{\gamma}{\nu}}, \quad \quad h_1(L) \sim L^{\frac{\Delta}{\nu}},
\label{102}
\end{equation}
which certainly holds below $d=d_c$ \cite{Ba83,Schladming}.
Equation (\ref{102}) also results from the Gaussian approximation. 
To distinguish the scaling forms (\ref{101}) and (\ref{102}), we refer to them  as $Q$-FSS and Gaussian FSS, respectively \cite{BeKe12}.  
$Q$-FSS is called modified FSS in reference \cite{BDT}, a term earlier used in reference \cite{KeLa93a} for the adjustments to standard FSS induced by  logarithmic corrections in four dimensions, following the introduction of a modified scaling variable in reference \cite{KeLa91}.
$Q$- or modified FSS has been verified many times over for PBC's in 4,5,6,7 and 8 dimensions  in references \cite{KeLa91,KeLa93,KeLa93a,Ke04,AkEr99,AkEr00,Ak01,MeEr04,MeBa05,AkEr01,MeDu06,JoYo05}.

However, for $\qq$ to be a new critical exponent of similar status to the others, it must be universal. 
For this to be the case, $Q$-FSS would have to hold independent of boundary conditions.
However, the standard belief for over 40 years is that systems with FBC's, in particular, have $\chi_L \sim L^2$ or $\qq=1$ \cite{RuGa85,LuMa11,Watson,Gunton}. The most recent independent numerical investigation of whether  Gaussian FSS or $Q$-FSS applies to the $d=5$ Ising model with FBC's, contained in reference \cite{LuMa11}, also supported the conventional belief that equation (\ref{102}) prevails.
If this were indeed the case, it would destroy the universality of  $\qq$ and the new hyperscaling relation (\ref{hhyperscaling}).
In reference \cite{BeKe12}, we examined the $d=5$ dimensional Ising model with PBC's and FBC's. 
We reaffirmed that $\qq=d/4$ in the PBC case and we provided overwhelming evidence that $\qq=d/4$ in the FBC case too.
Here we focus on the latter result, since that is the one most critical for $Q$-theory and the most at odds with standard opinion in the literature.

\section{Numerical evidence for universality}
\label{Numerical}

\begin{figure}[t]
\centerline{\includegraphics[width=0.55\textwidth]{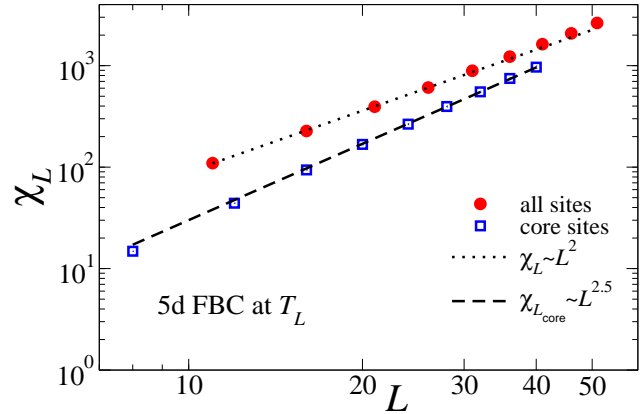}}
\caption{FSS for the pseudocritical Ising susceptibility calculated on 5D lattices using all sites (upper data set, in red) and using only core sites (lower data, in blue). The upper data set appears to scale as $\chi_L \sim L^2$, indicative of Gaussian FSS, but this is spurious and due to the preponderence of surface sites. The lower data, which are genuinely five-dimensional, scale according to the $Q$-FSS form $\chi_L \sim L^{5/2}$, supporting the universality of $\qq$.
(See also figure~4(a) of reference~\cite{BeKe12}.) } 
\label{fig1}
\end{figure}

In this section, we provide evidence for the universality of $\qq$.
The evidence we present is that $Q$-FSS holds for FBC's, just as it does for PBC's for the Ising model in five dimensions.

As stated, it is well established that, above the upper critical dimension, the Ising model on PBC lattices obeys $Q$-FSS at the critical point.  
In particular, the formula (\ref{101}) has been verified many times for the susceptibility at criticality \cite{AkEr99,AkEr00,MeEr04,MeBa05,AkEr01,MeDu06,JoYo05}.
The corresponding expression for the Lee-Yang zeros has been verified in reference \cite{BeKe12}.
It will come as no great surprise to the reader to know that the same forms govern FSS at the pseudocritical point in the PBC case too. This has also been verified for susceptibility and the zeros in reference \cite{BeKe12}. (In fact, $Q$-FSS has also been verified for the magnetization as well, both at the critical and pseudocritical points in reference \cite{BeKe12}.)

Long-standing belief is that the $Q$-FSS form (\ref{101}) {\emph{does not hold}} for FBC's above the upper critical dimension. Instead, standard belief is that the Gaussian form (\ref{102}), with $\qq=1$  holds there \cite{RuGa85,LuMa11,Watson,Gunton}.
Indeed, recent, explicit numerical support for $\chi_L \sim L^{\gamma / \nu} = L^2$ at the critical point was given in reference \cite{LuMa11}.
We contend that this interpretation is incorrect.
Our proposition is that $Q$-FSS applies in the FBC case above $d=4$, just as it does in the PBC case.
However, to observe it, one must perform FSS for FBC's at the pseudocritical point, not the critical one.
We will demonstrate that the infinite-volume critical point is too far away from the pseudocritical point to ``feel'' the finite-size scaling regime.

For the Ising model, upon which our numerical evidence is based, 
the partition function is
\begin{equation}
 Z_L(t,h) = \sum_{\{s_i\}}{
                           \exp{
                                 \left({
                                        -\beta E -hM
                                 }\right)
                          }},
 \label{Z}
\end{equation}
in which $E=-\sum_{\langle{i,j}\rangle}{s_is_j}$, $M=\sum_i{s_i}$,
$\beta = 1/kT$, $h=\beta H$, $H$ is the strength of an external field, $k$ is Boltzmann's constant and $s_i$ represents the value of an Ising spin at site $i$ of the lattice.
The summation in equation (\ref{Z}) is over spin configurations and the sum 
over $\langle{i,j}\rangle$ is over nearest-neighbour pairs of sites on the hypercubic lattice.
For a finite-size system, we measure the overall  susceptibility as 
\begin{equation}
 \chi_L(t) = L^{-d}\left({
  \langle{M^2}\rangle - \langle{|M|}\rangle^2
  }\right).
\label{chiL}
\end{equation}

\begin{figure}[t]
\centerline{\includegraphics[width=0.55\textwidth]{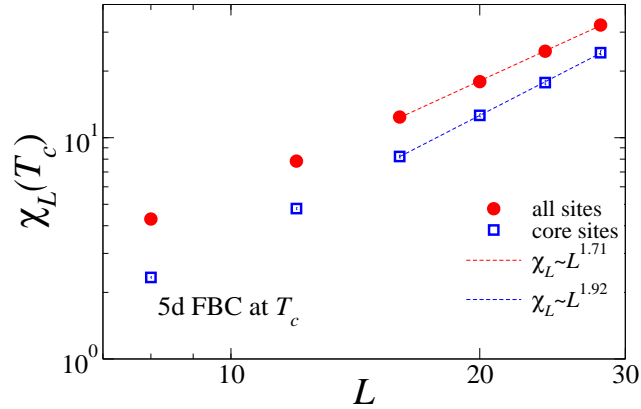}}
\caption{FSS for the 5D Ising susceptibility at the critical point $T_c$ from all sites (upper red data) and core sites (lower blue data). 
Neither Gaussian nor $Q$-FSS is supported. (See also figure~5(a) of reference~\cite{BeKe12}.)
} 
\label{fig2}
\end{figure}

For a size-$L$ hypercubic lattice, with FBC's, only $(L-2)^d$ sites are in the interior or bulk. 
Spins located on these interior sites interact with $2d$ nearest neighbours.
In this sense, they are genuinely immersed in a $d$-dimensional medium. 
The remaining $L^d - (L-2)^d$ spins are located on a surface of dimensionality $d-1$ or lower and interact directly with correspondingly fewer neighbours.
These are not, therefore, fully immersed in $d$-dimensions.
For example, with $L=24$ sites in each direction, the largest lattices analysed in reference \cite{LuMa11} had only 65\% of sites in the bulk. 
Therefore, the resulting calculations of $\chi_L$ are not truly representative of five dimensionality. Obviously the smaller lattices are even less representative of 5D.

To get around this problem and truly probe the five-dimensionality of the FBC lattices, we decided to remove the contributions of the outer layers of sites to equation (\ref{chiL}) and to other observables. 
We implemented this by omitting the contributions of the $L/4$ sites near each boundary and keep only the contributions of the $(L/2)^d$ interior sites. Each of these sites has 10 neighbours, as it should in five dimensions.

FSS for the susceptibility in the FBC case is represented in figure~\ref{fig1} where $\chi_L$ is plotted against $L$ on a logarithmic scale at pseudocriticality. 
The upper data set (red circles) corresponds to calculations of the susceptibility $\chi_L$ in equation (\ref{chiL}) using all lattice sites. 
The dotted line is a best fit to the form $\chi_L \sim L^2$. 
At first sight this appears to be a rather good fit,  supportive of the Gaussian FSS formula (\ref{102}) with $\gamma / \nu = 2$ and of the traditional view of FSS for FBC lattices.
However, closer inspection shows some deviation of the large-$L$ data from the dotted line. 
We interpret this as signaling that the apparently good fit to $\chi_L \sim L^2$ is spurious.

\begin{figure}[t]
\centerline{\includegraphics[width=0.55\textwidth]{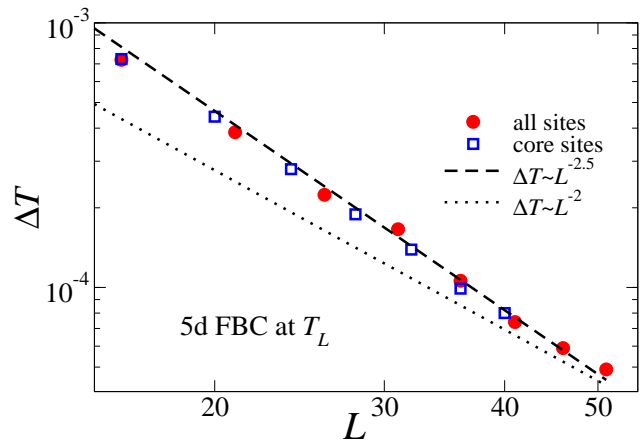}}
\caption{FSS of the susceptibility rounding is compatible with the $Q$-FSS $\theta  = \qq /\nu = 5/2$ and not with the Gaussian FSS $\theta  = 1/\nu = 2$.} 
\label{fig3}
\end{figure}

The lower (blue) data set in figure~\ref{fig1} corresponds to calculations of $\chi_L$ using the interior $L/2$ lattice sites only. The dashed line is a best fit to the $Q$-FSS form (\ref{101}) with $\qq \gamma / \nu = 5/2$ and the line clearly describes the large-$L$ data well.
This is evidence that the Ising model defined on the five-dimensional core of the  $L^5$ lattices obeys $Q$-FSS (\ref{101}) rather than Gaussian FSS (\ref{102}). This, in turn, is evidence for the universality of $\qq$.

In figure~\ref{fig2}, we present FSS for the susceptibility at the critical point $T_c$ with FBC's.
Again, the upper data set represents the circumstance where all sites contribute to (\ref{chiL}) and the lower data set corresponds to the usage of core sites only. 
Neither set of data is compatible with Gaussian or $Q$-FSS.

Thus $Q$-FSS applies at the pseudocritical point in the $d=5$ Ising model with FBC's, but neither 
$Q$-FSS nor Gaussian FSS applies at the critical point.
To investigate why, we next examine the rounding and shifting.
Both of these arise in any finite-size system because the counterpart of the divergence at $T_c$ in infinite volume is a smoothened peak. The width of the peak  called its rounding and the location of the peak (the pseudocritical point) is shifted with respect to the critical point.
If the rounding, $\Delta T$, is defined as the width of the susceptibility curve at half of its peak height (the ``half-height width''), then  one expects 
\begin{equation}
 \Delta T \sim L^{-\theta},
\end{equation}
where $\theta$ is called the rounding exponent. 
If $t_L = |T_L - T_c|/T_c$ is the  shift of the pseudocritical point relative to the critical one, then one has 
\begin{equation}
 t_L \sim L^{-\lambda},
\end{equation}
where $\lambda$ is the shift exponent. 
For standard ($d<d_c$) FSS, one normally has $\theta = \lambda = 1/\nu$ although this is not always the case. Above the upper critical dimension this would lead to $\theta = 2$.
For $Q$-FSS, one may expect that $\theta = \lambda = \qq /\nu = 5/2$, but, again, this is not a requirement.
Our main concern is the relative sizes of the rounding and the shifting in the FBC case.
If the shifting  is bigger than the rounding then the infinite-volume critical point $T_c$ will be too far away from the pseudocritical point to come under its influence -- it will be outside the FSS regime.
This would explain why FSS at the critical point is different to FSS at the pseudocritical point.
In this case, FSS at $T_c$ would certainly not be $Q$-FSS, but there is no reason for it to be Gaussian FSS either.

\begin{figure}[t]
\centerline{\includegraphics[width=0.55\textwidth]{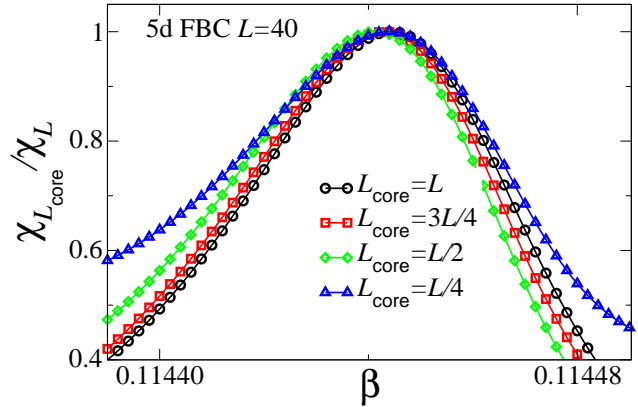}}
\caption{The value of the pseudocritical temperature is independent of whether it is defined from the susceptibility peak using all lattice sites or only the core.
(See  figure~7 of reference~\cite{BeKe12}.) } 
\label{fig4}
\end{figure}

The rounding is investigated in figure~\ref{fig3} for the entire lattice and its core.
It is obvious that the rounding is not of the standard, Gaussian type.
Instead, it follows the $Q$-theoretic expectation, with rounding exponent $\theta  = \qq /\nu = 5/2$.
This means that the rounding is sharper than one may naively expect, or the FSS regime is narrower than what the Gaussian theory would deliver.

In figure~\ref{fig4} we plot the ratio of the core susceptibility to the total susceptibility for the FBC case and for various definitions of the core. 
Whether the susceptibility is determined using the innermost 25\%, 50\% or 75\% of  sites in each direction, or, indeed, whether it comprises contributions from all sites including the boundaries, makes little difference to the location of the susceptibility peak.  
The shift exponent is measured in figure~\ref{fig5} using contributions to the susceptibility from the entire lattice.    The shift exponent is clearly $\lambda = 1/\nu = 2$.

Thus the shifting is indeed bigger than the rounding. 
Therefore the critical point $T_c$ is too far away from the pseudocritical point $T_L$ to feel its influence.
In other words, the finite-size susceptibility at the critical point is outside the pseudocritical FSS domain. This explains the results in figure~\ref{fig2} 
- the remoteness of $T_c$ from $T_L$ means the plot is beyond the  FSS regime.

To complete our investigations of equations (\ref{101}) and (\ref{102}) in the FBC case, the FSS for the Lee-Yang zeros is tested in figure~\ref{fig6} at the pseudocritical point. 
In fact we present the scaling of the first two zeros for FBC lattices using the contributions from all sites and from the core-lattice sites only.
In each case the zeros scale with as $L^{-q\Delta / \nu} = L^{-15/4}$ according to the $Q$-FSS formula on the right of equation~(\ref{101}) rather than the traditional, Gaussian formula in equation~(\ref{102}).

\section{Dangerous irrelevant variables}

The origins of the new exponent $\qq$ can be explained through the  dangerous irrelevant mechanism in the renormalization group.  
Standard FSS in $d < d_c$ may be understood by writing the finite-size free energy below the upper critical dimension as \cite{PrFi84} 
\begin{equation}
 f_L(t,h) = b^{-d} f_{L/b}
 \left({
  tb^{y_t},hb^{y_h}
 }\right).
\label{woDIV}
\end{equation}
The  correlation length is
\begin{equation}
\xi_L(t,h) = b \Xi_{L/b} \left({tb^{y_t},h b^{y_h}}\right),
\label{BNPY200}
\end{equation}
which identifies $\nu = 1/y_t$ through setting $b=L$ and $h=0$, and then taking the limit $L \rightarrow \infty$. At $t=0$, it gives $\xi_L \sim L $.
In the absence of the external field $h$, equation (\ref{woDIV}) can be written 
$ f_L(t,0) = L^{-d} f_{1} \left({(L/\xi_\infty )^{1/\nu},0 }\right) $.
Thus standard FSS is controlled by the ratio $x=L/\xi_\infty(t)$. 
Moreover, in the scaling regime where $x \ll 1$, the function
 $f_1(x^{1/\nu})\sim x^d$ is universal, leading to $f_L(t)\sim t^{d\nu}$.
 Differentiating this twice delivers the specific heat and standard hyperscaling (\ref{hyperscaling}).

\begin{figure}[t]
\centerline{\includegraphics[width=0.55\textwidth]{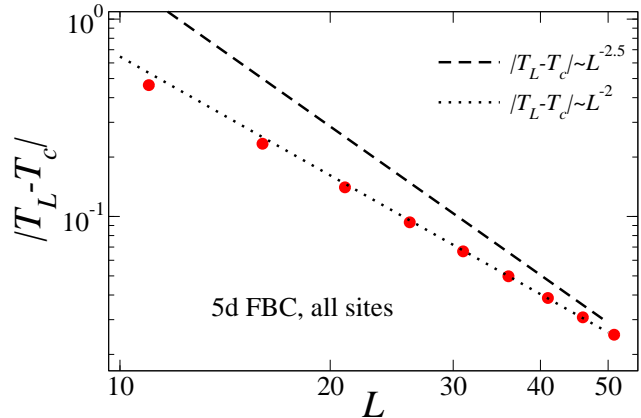}}
\caption{The high-dimensional Ising shift exponent $\lambda$ for FBC's comes from the Gaussian formula $\lambda = 1/\nu = 2$. This means that the shifting is bigger than the rounding in the FBC case, and the critical point $T_c$ is too far from the pseudocritical point $T_L$ to come under its scaling influence.
(See also figure~6(b) of reference~\cite{BeKe12}.) } 
\label{fig5}
\end{figure}

A more complete version of equation~(\ref{woDIV}) is \cite{FiHa83}
\begin{equation}
f_L(t,h,u) = b^{-d} f_{L/b} \left({tb^{y_t},hL^{y_h},uL^{y_u}}\right).
\label{BL4.2} 
\end{equation}
Here $u$ is associated with the $\phi^4$ term in the Ginzburg-Landau-Wilson action.
For $d<d_c$, it is a relevant scaling field, but at  $d_c$, $u$ becomes marginal and above it  is  irrelevant. 
 There the Gaussian fixed point controls critical behaviour with
$y_t = 2$, $y_h = 1+{d}/{2}$ and $y_u = 4-d$  \cite{Ma}.
Naively  differentiating equation (\ref{woDIV}) or (\ref{BL4.2}) delivers  functions different to those from Landau theory.
This is because the limit $u\to 0$ is singular, and has to be properly  accounted for.
For this reason, $u$ is termed a dangerous irrelevant variable.
Its proper treatment leads to  rewriting equation (\ref{BL4.2}) as \cite{BNPY,BZJ85}
\begin{equation}
f_L(t,h,u) = b^{-d} f_{L/b} \left({tb^{y_t^*},hb^{y_h^*}}\right)
= L^{-d} f_{1} \left({tL^{y_t^*},hL^{y_h^*}}\right),
\label{BNPY}
\end{equation}
in which
\begin{equation}
y_t^* =y_t - \frac{y_u}{2}=\frac{d}{2}, \quad \quad 
y_h^* =y_h- \frac{y_u}{4} =\frac{3d}{4} .
\label{scytstar}
\end{equation}
Similar considerations for the correlation length deliver 
\begin{equation}
\xi_L(t,h,u) = L^{\sqq} \Xi \left({tL^{y_t^{*}},h L^{y_h^{*}}}\right).
\label{BNPY2}
\end{equation} 
In reference \cite{BNPY} the assumption was made that $\qq=1$.
This was driven by the belief that ``the correlation length $\xi_L$ is bounded by $L$'' even for PBC's. 
In this case a second length scale would be needed to modify FSS \cite{Bi85,Binder87}.
Introducing $\ell_\infty(t) \sim t^{-1/y_t^*}$, the first argument on the right-hand side of  equation~(\ref{BNPY}) or (\ref{BNPY2}) may be written $(\ell_\infty(t)/L)^{y_t^*}$
and $\ell_\infty(t)$ was deemed to control FSS.
It was dubbed the {\emph{thermodynamic length}} by Binder \cite{Bi85}.
Its finite-size counterpart  $\ell_L$ was called the {\emph{coherence length}}
in reference \cite{BDT}, where a so-called {\emph{characteristic length}} $\lambda_L(t)$ was also introduced as the FSS counterpart of the infinite-volume correlation length. 

From our considerations, it is clear that this plethora of different lengths is unnecessary;
the exponent  $\qq$ in equation (\ref{BNPY2}) is not bounded by $1$.
A direct, explicit, numerical calculation of the FSS of the correlation length for the 5D PBC model in reference \cite{JoYo05} showed $\xi_L \sim L^{5/4}$ there. 
This was verified in reference \cite{BeKe12}.
It is by now well established that the replacement of the scaling variable $L/\xi_\infty(t)$ of standard FSS by $L^{\sqq}/\xi_\infty(t)$ of $Q$-FSS is correct for the susceptibility, magnetization and pseudocritical point in periodic Ising models in four \cite{JoYo05,KeLa91}, five \cite{RiNi94,PaRu96,JoYo05,AkEr99}, six \cite{AkEr00,MeEr04,MeBa05}, seven \cite{AkEr01} and eight \cite{MeDu06} dimensions. The results presented here and in reference \cite{BeKe12} support our assertion that the same holds true for FBC's and that $\qq$ is universal.

Thus the breakdown in standard hyperscaling (\ref{hyperscaling}) above the upper critical dimension may be explained through dangerous irrelevant variables. 
The breakdown of FSS was less clear, however. Although the above formulation in terms of dangerous irrelevant variables does not involve explicit statements about boundary conditions, while it has  been broadly accepted for PBC's, Gaussian FSS was believed to hold in the FBC case. In this sense, 
standard FSS was not universal after all, a circumstance which was ``poorly understood'' \cite{JoYo05,ChDo}.

We have now shown that FSS for FBC's is the same as for PBC's at pseudocriticality, but not at criticality and this is associated with  the universality of the new exponent $\qq$. 
However, the logarithmic counterpart to $\qq$ cannot be attributable to dangerous irrelvant variables, since these arise only for $d>d_c$ and non-trivial $\hat{\qq}$ necessitates $d=d_c$.
The reader is referred to reference \cite{KJJ2006} for this circumstance.

\begin{figure}[t]
\centerline{\includegraphics[width=0.55\textwidth]{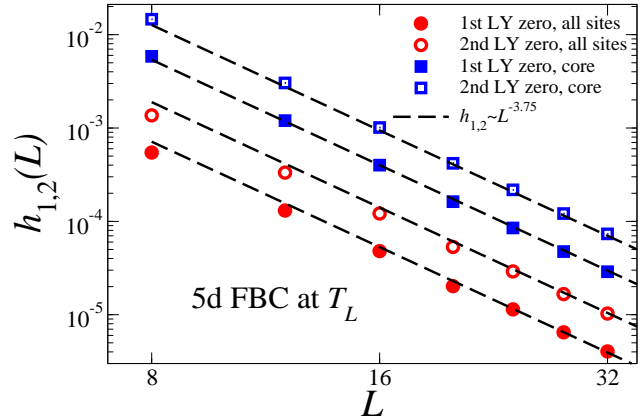}}
\caption{The first two Lee-Yang zeros for Ising systems with FBC's at pseudocriticality obey $Q$-FSS whether the full lattice or only the core sites are used in their determination.
(See also figure~8(c) of reference~\cite{BeKe12}.) } 
\label{fig6}
\end{figure}

\section{Conclusions}

It is well known that standard FSS is universal below the upper critical dimension $d=d_c$ when hyperscaling holds and where the correlation length is comparable to the actual extent $L$ of a system.
Above $d_c$, the breakdown of  standard  hyperscaling  is attributed to dangerous irrelevant variables  in the renormalization-group  approach.
Although closely related to hyperscaling, FSS was until now believed to be non-universal in high dimensions, with equation (\ref{101}) holding for FBC's and (\ref{102}) for PBC's.
Although this picture appeared to be supported numerically for FBC's in references~\cite{RuGa85,LuMa11} 
and for PBC's in references~\cite{KeLa91,KeLa93,KeLa93a,Ke04,AkEr99,AkEr00,Ak01,MeEr04,MeBa05,AkEr01,MeDu06,JoYo05},
it was unexplained why the dangerous irrelevant variable mechanism
should apply in the one case and not in the other. 

Here we have used Lee-Yang zeros to show that the scaling mechanism is self consistent only if the correlation length scales as a power of the length above $d_c$.
This is the case irrespective of boundary conditions and leads to the  introduction of a new scaling exponent, which we denote by $\qq$. 
Since it is universal, $\qq$ has a similar status to the critical exponents  $\alpha$, $\beta$, $\gamma$, $\delta$, $\eta$ and $\nu$, in notation standardised by Fisher in the 1960's.
The introduction of $\qq$ allows one to extend the dangerous-irrelevant-variable mechanism to the correlation length through equation (\ref{BNPY2}).
FSS is then implemented by the substitution $t \rightarrow L^{-\sqq/\nu}$, a procedure we term ``$Q$-FSS'' to distinguish it from the standard  $t \rightarrow L^{-1/\nu}$ valid below $d_c$.

Here we point out that, for the FBC lattice sizes used in reference \cite{LuMa11}, the bulk of sites are on the surface, so that the system is not genuinely five-dimensional. 
The resulting  conclusion that $\chi_L$ obeys Gaussian FSS is  not a 5D one.
For this reason, we re-examined FSS for the 5D Ising model.
In order to probe the five-dimensionality of the structure, 
we remove contributions close to the lattice boundary. 
In addition to FSS at the critical temperature, we also examined pseudocriticality.
Our numerical results indicate that, once the lower-dimensional influence of the peripheries is removed, the FBC lattice exhibits the {\emph{same}} scaling as the PBC one at pseudocriticality,
namely that given by $Q$-FSS.
Using the same technique at the infinite volume critical point, we find no evidence for either Gaussian FSS or for $Q$-FSS.
Because the rounding is smaller than the shifting, we attribute this to the fact that $T_c$ is too far from $T_L$ to come under the influence of FSS there. 
This means that the conventional  FSS paradigm for FBC lattices above $d_c$  is {\emph{unsupported}}, and this is particularly clear at pseudocriticality \cite{RuGa85,LuMa11,Watson,Gunton}.
It also offers evidence for the  universality of $\qq$ at pseudocriticality and introduces a new, 
universal version of hyperscaling through equation (\ref{hhyperscaling}), which is valid in all dimensions.

\section*{Acknowledgements}

We are  grateful to J.-C.~Walter who performed the simulations.
We also thank M.E.~Fisher for a number of helpful suggestions including to introduce the symbol $\qq$ for the new exponent \cite{coppa}. 
This work is supported by the EU Programmes FP7-People-2010-IRSES (Project
No. 269139) and FP7-People-2011-IRSES (Project No. 295302).


%
%

\end{document}